%% file: berggren-sgv-LCWS2011granada-proceedings-sub.tex
\documentclass[twoside]{LCWS11}
\usepackage[latin1]{inputenc}
\usepackage[dvips]{graphicx,epsfig,color}
\usepackage{wrapfig,rotating}
\usepackage{amssymb,amsmath,array}
\usepackage{enumitem}
\usepackage{cite}

\input{newcom}

\begin{document}
\title{
SGV 3.0 - a fast detector simulation}
%***********************************************************************
% AUTHORS INFORMATION AREA
%***********************************************************************
\author{Mikael Berggren
% Optional short acknowledgment: remove next line if non-needed
% DO NOT MODIFY THE FOLLOWING '\vspace' ARGUMENT
\vspace{.3cm}\\
% Addresses and institutions (remove "1- " in case of a single institution)
DESY \\
Notkestra\ss e 85, D-22607 Hamburg -Germany
}
%%***********************************************************************
% END OF AUTHORS INFORMATION AREA
%***********************************************************************

\maketitle
\begin{abstract}
The need for fast simulation programs is emphasised, both
in terms of the need for ``rapid response'' to new
results - in particular from the LHC - and new theoretical ideas,
and in terms of how to cope with  multi-billion simulated event samples.
The latter would arise both from the need to be able to
simulate significantly more events than expected in the real data, also for
high cross-section processes, and the need to scan multi-parameter
theories.

The {\it Simulation \`a Grande Vitesse}, SGV, is presented, and is shown to
be able to address these issues. 
The tracking performance of SGV is shown to reproduce
very closely that of the full simulation and reconstruction of the ILD
concept. Preliminary results on how to also closely emulate the
calorimetric performance from full simulation is presented.
The procedure is parametric, with no
need to simulate the detailed shower development, and promises
to be many orders of magnitude faster than such approaches. Contrary to
what is often the case with fast simulation programs, the procedure
gives a somewhat {\it pessimistic} result, compared to the full simulation
and reconstruction.
\end{abstract}

\section{Introduction}\label{Sect:introduction}
The latest years of development has brought forward
very performant and complete full simulation packages, both
in SiD and ILD~\cite{Bib:LOIild,Bib:LOIsid}.
One might then wonder why there still is any need for fast
simulation of the ILC or CLIC detectors.

One answer to this was given during this conference by R. Heuer,
when he pointed out that ``{\it We need to update the physics case} (for the LC) 
{\it continuously}''.
This means that not only does one need detailed simulation of
a few bench-mark reactions - to validate the detector concepts -
but also to simulate a large variety of processes, possibly
from newly conceived models, or models that recently have been
revised in light of observations at the LHC.
To this can be added that the studies leading to the LOI~\cite{Bib:LOIild,Bib:LOIsid} showed that, 
as far as physics
results were concerned,
fast and full simulation studies gave close to identical results.
To fulfil the needs for fast and precise physics results, 
such fast simulation programs need to be  light-weight, to be able to run 
without the need of large computer resources, with a low threshold for non-experts
to start using it,
but most of all they need to be truly fast.

There are two cases where the speed of the simulation is of
utmost importance: High cross-section background process, and
multi-parameter theory scans.
\subsection{Examples: $\gamma\gamma$ cross-sections and SUSY scans}\label{Ssect:examples}

At $\sqrt{s}$ = 500 \GeV, PYTHIA~\cite{Bib:pythia} estimates the total cross-section 
for $e^{+}e^{-}\rightarrow\gamma\gamma e^{+}e^{-}\rightarrow q\bar{q}e^{+}e^{-}$
at $E_{CMS}=500$ GeV to be 35 000 pb.
Therefore, 17.5 $\times$$10^{9}$ such events would have been produced
after taking 500 fb$^{-1}$ of data, which each ILC experiment expects
to have collected in the first four years of running.
A typical time to generate such events is 10 ms.
A fast detector simulation would aim at not requiring more than
that for the detector simulation.
This leads to a total time
of   3.5 $\times$$10^{8}$ s, equal to about 10 years, to generate
500 fb$^{-1}$ of  $\gamma\gamma$ events.
Already with only a handful of CPU's,
such a sample could be simulated in a few months, and with typical resource of 
a batch-farm (several hundred cores), not more than a few days
would be needed.
On the other hand, full simulation and reconstruction takes several
minutes per event, ie. more than three orders of magnitude more, and would require
many thousand CPU-years. Even with full-blown grid-processing,
such a program would take years, possibly longer than it would take to
collect the real data. It should also be noted that
these numbers apply for the modest requirement that the simulated sample
is the same size as the real data. This is arguably far from being
sufficient: with such a small sample, the simulation statistical error
might be the dominating systematic error of the measured quantities.
Probably one would require as an absolute minimum that the simulated
sample is five times larger than the data sample; this was the case at LEP.

An other example where very large samples need to be 
simulated is scanning SUSY parameter-space.
MSUGRA can serve as a simple example:
In this model, there are four continuously varying theory parameters and the sign of a fifth one.
If one wants to scan each of the continuous parameters in 20 steps,
and simulate 5000 events per point\footnote{A modest requirement: in eg. the MSUGRA point 
sps1a' almost
1 million SUSY events are 
expected for 500 fb$^{-1}$.}, about 2$\times 10^9$ events need to be
simulated.
Such events are slower to generate and simulate than $\gamma\gamma$ events,
so also in this case  CPU millennia would be needed to do full simulation.

\section{The SGV fast simulation}\label{Sec:SGV}

Fast detector simulations exist of different types, with 
different levels of sophistication. 
For any approach, the aim is that the
detector-simulation time of one event should be of the same order
as the time to generate an event
by an efficient event generator, such as PYTHIA6.
One can make
a simple smearing of the generated four-vectors 
with some global assumed detector
properties. A somewhat more elaborate scheme is the
parametric simulation, where measurements are parametrised
with respect to particle energy and angle. SIMDET, the fast simulation
program used for the TESLA TDR physics studies falls in this
category~\cite{Bib:tesla}. Finally,
one can construct covariance matrix machines,
where the full covariance matrix is constructed from
the generated particles and the detector layout.
In this category one finds eg. LiCToy~\cite{Bib:lictoy} and SGV - {\it La Simulation \`a Grande
Vitesse}.

SGV was originally developed in the early nineties as a tool to evaluate the proposed
upgrade of the DELPHI detector in view of the new conditions expected
due to the transition form LEP I to LEP II~\cite{Bib:delphivft}. It evolved into a valuable tool
for new physics searches in DELPHI, both for signal and background simulation~\cite{Bib:delphi_mssm}.
It has subsequently been used for physics and detector studies for TESLA,
LDC and ILD~\cite{Bib:tesla,Bib:ILC_det,Bib:LOIild,Bib:ILC_phys}.

Over the last year, the well-tested SGV2 series (written in Fortran77) have been rewritten and
re-organised into an SVN-managed Fortran95 package. Most of the
previous versions dependence on CERNLIB has been removed\footnote{Work will be done to 
further reduce CERNLIB dependence.
This will inevitably be at a the cost of
backward compatibility on steering files, as the usage of the FFREAD
package would in that case be replaced by using Fortran namelists, which
has the same functionality, but different syntax. HBOOK dependence 
will remain in the foreseeable future - but only for user convenience :
SGV itself doesn't need it.},
the installation procedure has been re-written, and new features have been added,
and more are planned.
SGV has been tested to work on both 32 and 64 bit architectures out-of-the-box,
and it was verified that the transformation from Fortran77 to Fortran95 did not
deteriorate the speed. In fact, the Fortran95 version was
found to be faster (by 15\%) than the Fortran77 version.

Among the features of SGV are:
    \begin{itemize}[noitemsep,label=$-$]
      \item Both  PYTHIA~\cite{Bib:pythia} and Whizard~\cite{Bib:whizard} are internally callable.
      \item Alternatively, input can be read from PYJETS or StdHep~\cite{Bib:stdhep}.
      \item The same formats can be used to output the generated event.
      \item A {\tt samples} subdirectory with steering and code for eg. scanning single
            particles, create an HBOOK ntuple with ``all'' information, which can be
            converted to ROOT using the  h2root tool. There is also code to
            output the simulated event in LCIO DST-format~\cite{Bib:lcio}.
    \end{itemize}
Features that are foreseen to be added to SGV in the near future are:
    \begin{itemize}[noitemsep,label=$-$]
      \item Development on calorimeters, which will be detailed in Chapter \ref{Sec:calotune}.
      \item Including a filter-mode, which would allow
to simulate large samples with varying detail as needed for
a specific analysis. One would
generate the event inside SGV and
subsequently run the SGV detector simulation and analysis. From the result
of the analysis, the fate of the event can be determined, from completely discarding it,
over filling control-histograms, writing it to an ntuple, to LCIO, or to request
full simulation, by outputting the generated event in  StdHep format.
    \end{itemize}

To install SGV, one should first execute (preferably in a new, empty directory):
\begin{verbatim}
svn export https://svnsrv.desy.de/public/sgv/tags/SGV-3.0rc1/ SGV-3.0rc1/
\end{verbatim}
followed by 
\begin{verbatim}
cd  SGV-3.0rc1 ; bash install
\end{verbatim}
These two commands will take about a minute to complete.
The main documentation is in the README file in the top-directory,
and for various specific tasks and examples, the README:s in
the {\tt samples} sub-directory and it's sub-directories. This allows
to get various external programs installed, if they are not
already available on the system. This includes
StdHep, CERNLIB (in native 64bit), Whizard (both basic and ILC-tuned versions),
and the LCIO-DST writer.

\subsection{Simulation of the tracking detectors}\label{Ssec:tracking}
\begin{wrapfigure}{r}{0.5\columnwidth}
       \includegraphics[scale=0.3]{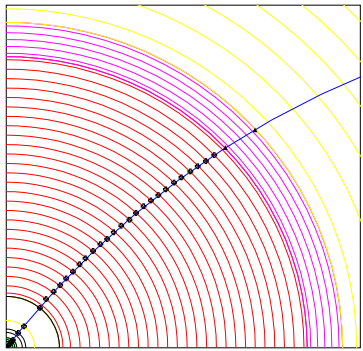}
\caption{A graphical rendering of the method described in the text.}\label{Fig:atrack}
\end{wrapfigure}

As stated above, SGV is a machine to calculate covariance matrices.
The procedure used is as follows~\cite{Bib:billoir}:
The track-helix is followed through the detector, to find what layers are 
hit by the particle, as illustrated in Figure \ref{Fig:atrack},
showing the $R\phi$ projection of a quadrant of the ILD detector,
as it is described in SGV. 
The outward tracking continues until the intersection of
the start of the out-most calorimeter is reached.  
The helix is locally described either by barrel coordinates,
or forward coordinates, depending of the nature of the intersected
surface. In the forward-barrel overlap region, it can possibly switch 
between these descriptions 
several times along it's trajectory.

From the list of intersected surfaces, the covariance matrix at the perigee is calculated:
The helix is followed from the outside, starting at the outer-most tracking-detector 
surface. At each recorded intersection, the measurements
the surface contributes are added in quadrature to the relevant elements of the
covariance matrix. The matrix is then inverted, to obtain the weight matrix.
The effect of multiple-scattering~\cite{Bib:lynchetdahl} at the surface is added to the relevant
elements of this matrix. The matrix is then once again inverted, and translated
along the helix (in five-dimensional helix-space) to the next
intersected surface, and the procedure is repeated. This continues until the mathematical
surface representing the point of closest approach is reached.
As each track is followed through the detector, the information on hit-pattern 
is automatically obtained, and is made accessible to later analysis. 

It can be noted that this method can be described in mathematical terms as
a realisation of a Kalman filter~\cite{Bib:kalman,Bib:billoir_kalman}, and often in particle physics
 ``Billoir track-fit'' and ``Kalman filter'' are treated as synonyms.
In SGV, the formalism of Kalman filters is not used, rather
all matrix operations, including the inversions, are worked out in element-form
in the code, to avoid having to call general-purpose numerical methods,
which possibly are inefficient for the problem at hand and might impede on the performance
of the optimisation done by the compiler. 

\begin{figure}[t]
      \includegraphics[scale=0.35]{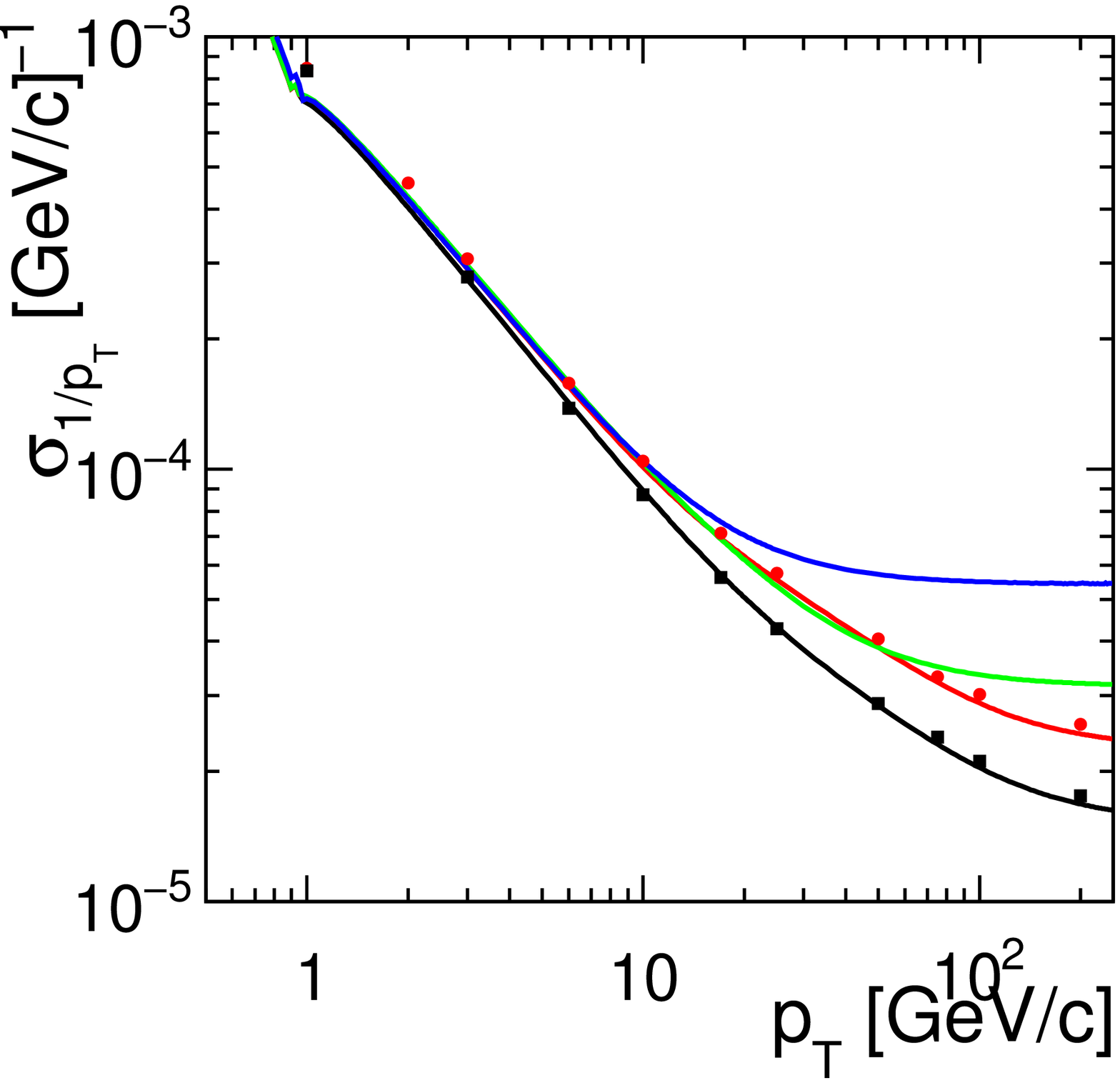}
       \includegraphics[scale=0.32]{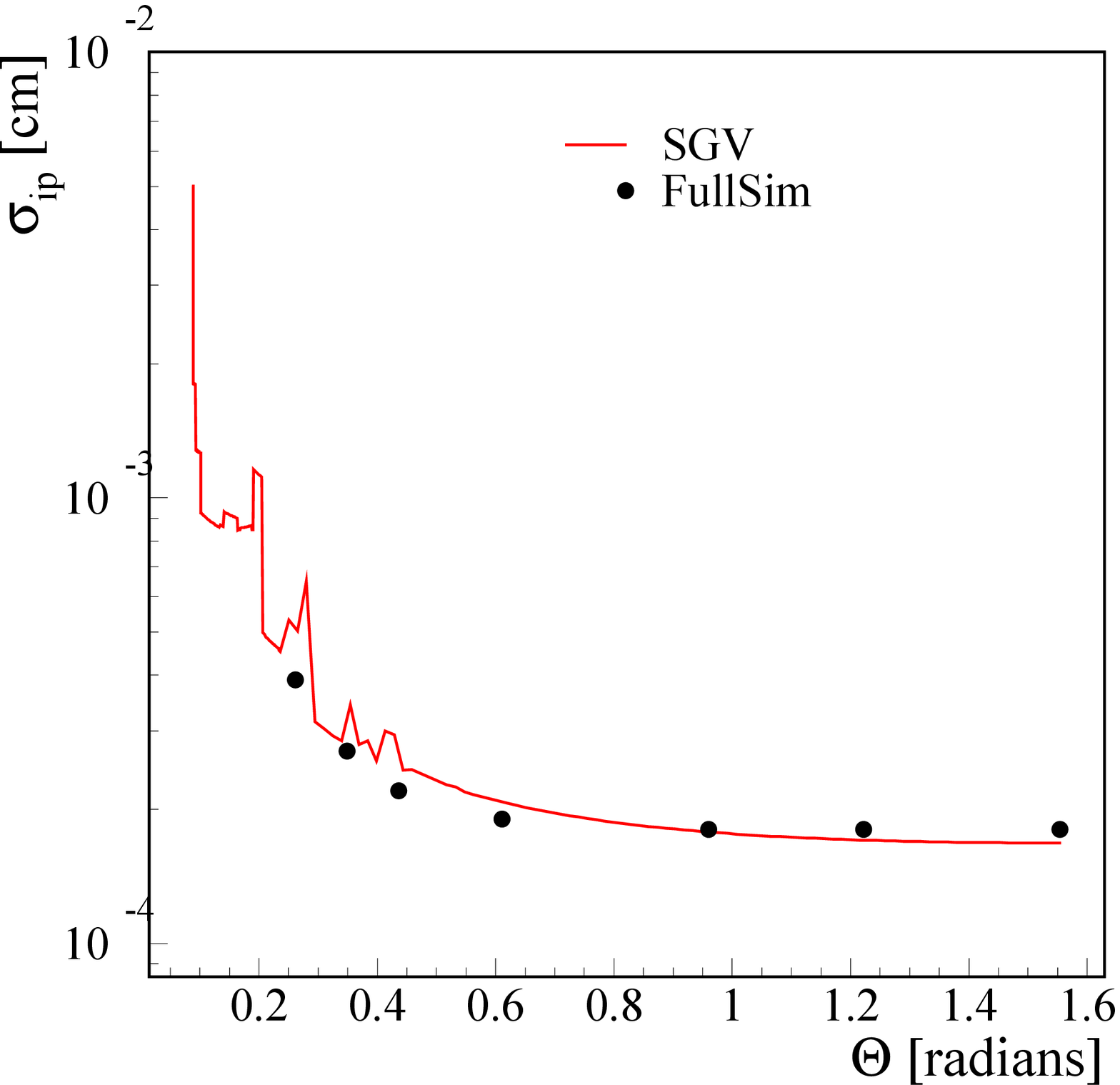}
\caption{Left plot: The momentum error $\sigma_{1/p_T}$ vs. $p_T$,
for a number of different detector configurations.
Right plot:  The impact-parameter error $\sigma_{ip}$ vs. $p_T$.
The  lines show the SGV result, and the dots show the full simulation
and reconstruction result.}\label{Fig:dp_vs_p_dipvsth}
\end{figure}

The perigee parameters are then smeared according to the calculated 
covariance matrix. This uses the method of doing a Cholesky decomposition~\cite{Bib:cholesky} 
of the matrix, and then multiplying the lower-triangular component ($L$) with a
vector ($u$) filled with uncorrelated random variables.
The product vector $v=Lu$ will contain random numbers with correlations between them that are
indeed those of the calculated covariance matrix. Figure \ref{Fig:dp_vs_p_dipvsth}
shows a few examples of the excellent agreement between the SGV result and that obtained by the full simulation and
reconstruction for the same detector configuration.
%Co-ordinates of hits accessible.

\subsection{Simulation of calorimeters}\label{Ssec:calosim_general}
To simulate calorimeters,
the charged or neutral particle is extrapolated to the intersections with the 
various calorimeters.
A decision is made on how the detectors will act. It can be concluded that the
particles should be detected as a minimum ionising one, or that it should
initiate an electromagnetic shower, or a hadronic shower, or 
that it is below the detectability threshold..
According to the chosen process, the detector response is simulated from parameters,
given in the geometry description input-file.
As a final (non-obligatory) step, showers can be merged if they are
sufficiently close.
The code that tracks the particles to the intersections is separated
from the code simulating the response, so by replacing the latter (at compile-time)
with a user-supplied routine, any other  shower-simulation can be used.
It should be kept in mind, however, that any more sophisticated algorithm
probably is orders of magnitude slower, and would denigrate the main
benefit of a fast simulation.
A step towards increasing the realism is to simulate confusion between
calorimetric clusters. The procedure to emulate the full reconstruction
in this respect 
is described below, in Section \ref{Sec:calotune}.
\subsection{Additional simulation features}\label{Ssec:added}
In addition to the above core-functionality of SGV, the program also
allows for the simulation of
electromagnetic interactions (pair-creation and bremsstrahlung)
in the detector material,
and has a well-defined scheme on how to plug in code simulating particle identification,
track-finding efficiencies (both on the whole-track and hit level),
and the presence of scintillators or taggers (ie. detector-elements that only
measure the presence, within the acceptance of the element, of particles above a threshold.)

\section{Calorimeter simulation tuning}\label{Sec:calotune}
The basic issues of a fast simulation of calorimeters - the random error on the detected energy, 
on the shower position, and on it's shape - are included by default in SGV.
However, there are  also association errors:
Clusters might merge, might split, or might get wrongly associated to tracks.
Since the measurement by the tracking system - if it is available - is 
always preferred to the calorimetric measurement,  association errors entails
errors on the total reconstructed energy:
On one hand, if a (part of) a 
neutral cluster gets associated to
a charged track,  energy is lost,
on the other hand,
if a (part of) a charged cluster is {\it not} 
associated to any track, energy is double-counted.
Other errors, eg. split neutral clusters, charged 
                cluster associated with wrong track and so on, are of
                less importance, since they do not give rise to an error on the total event energy or momentum. 

\begin{figure}[b]
      \includegraphics[scale=0.32]{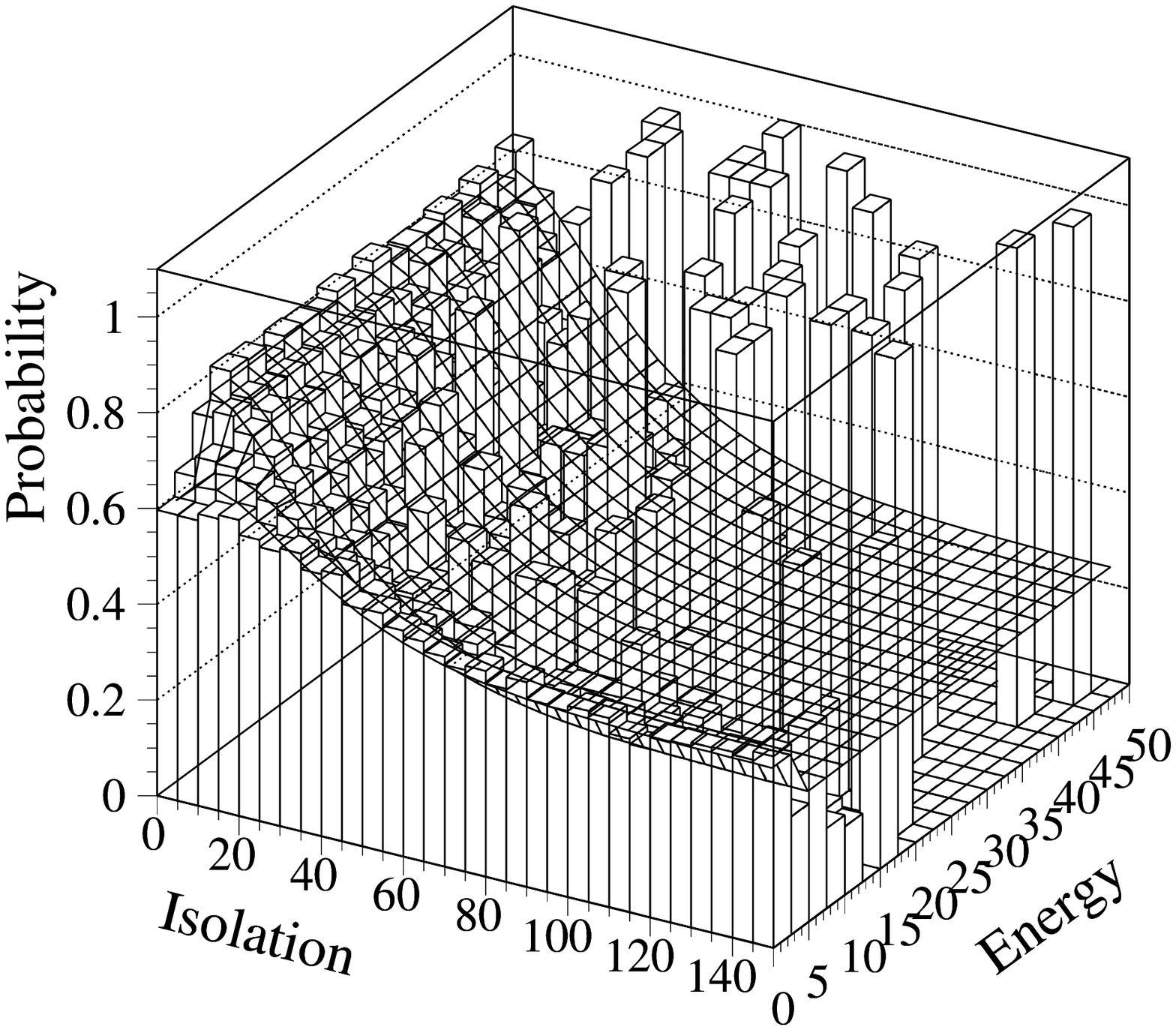}
      \includegraphics[scale=0.32]{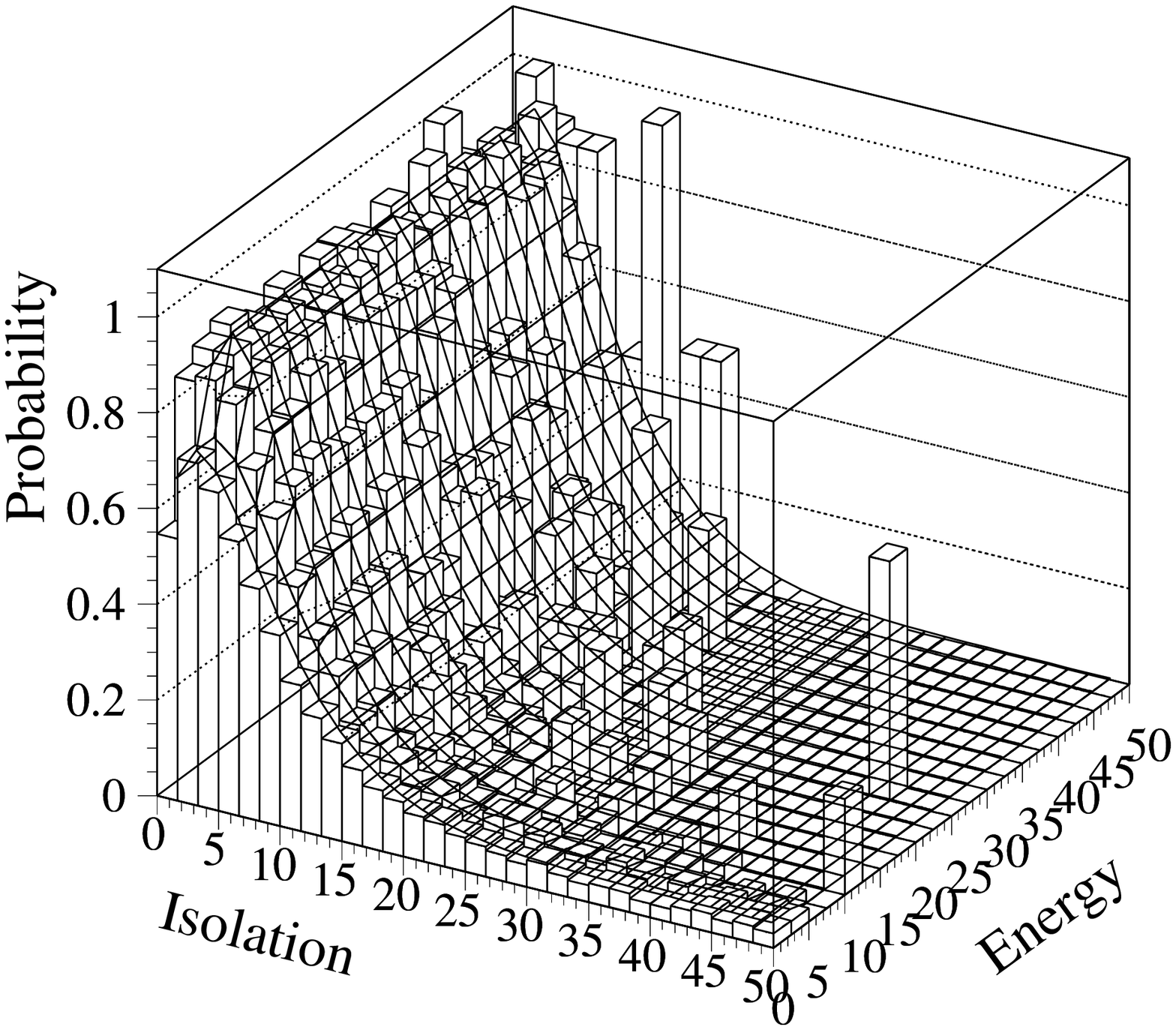}
\caption{The probability to split clusters versus energy and isolation. The left plot
 shows the situation for charged hadrons (double-counting),
while the right plots shows the situation for photons (energy loss).
The histograms shows the observed performance of Pandora, while the
mesh is the fit.
}\label{Fig:split}
\end{figure}
In SGV, information is already 
available about where the particle hits the calorimeters.
The program contains procedures to generate errors on energy, position and 
              shower-axes from geometry file input parameters,
to merge clusters based on generated
              shower positions and axes to
accommodate errors in the association between clusters and tracks. 
All these procedures can be controlled by the SGV geometry and steering-files.
Therefore, the next step to further increase the realism of the simulation is
to treat the association errors.

To study  association errors, a sample was selected from the LOI 
mass-production - 8 thousand  $\eeto udsc$ fully simulated and reconstructed events.
The particles reconstructed using the particle-flow algorithm Pandora~\cite{Bib:pandora} were 
compared to the true particles.
To be able to compare only the effects of the treatment of calorimeters, not differences
in the treatment of interactions and measurement in the tracking volume, the true particles and
reconstructed tracks were read from the full simulation DST.
The calorimeter hits were also read from there, and were used to create true clusters, ie.
clusters made exclusively by calorimeter hits created by a certain true particle.
The study concentrated on the most important issues, ie. double-counting and energy loss,
while neutral-neutral or charged-charged merging was not considered,
nor was multiple splitting/merging.
Among the observables  available in fast simulation, the most relevant ones were then identified.
This included the cluster energy,
the distance at the calorimeter face to nearest true particle of ``the other type'' (ie. neutral-to-charged 
or charged-to-neutral),
whether the particle was a hadron or not, and whether it would be detected by
the barrel or end-cap calorimeters.
The confusion was then broken down into sub-processes and was found to be
possible to factorise as:
\begin{figure}[t]
      \includegraphics[scale=0.32]{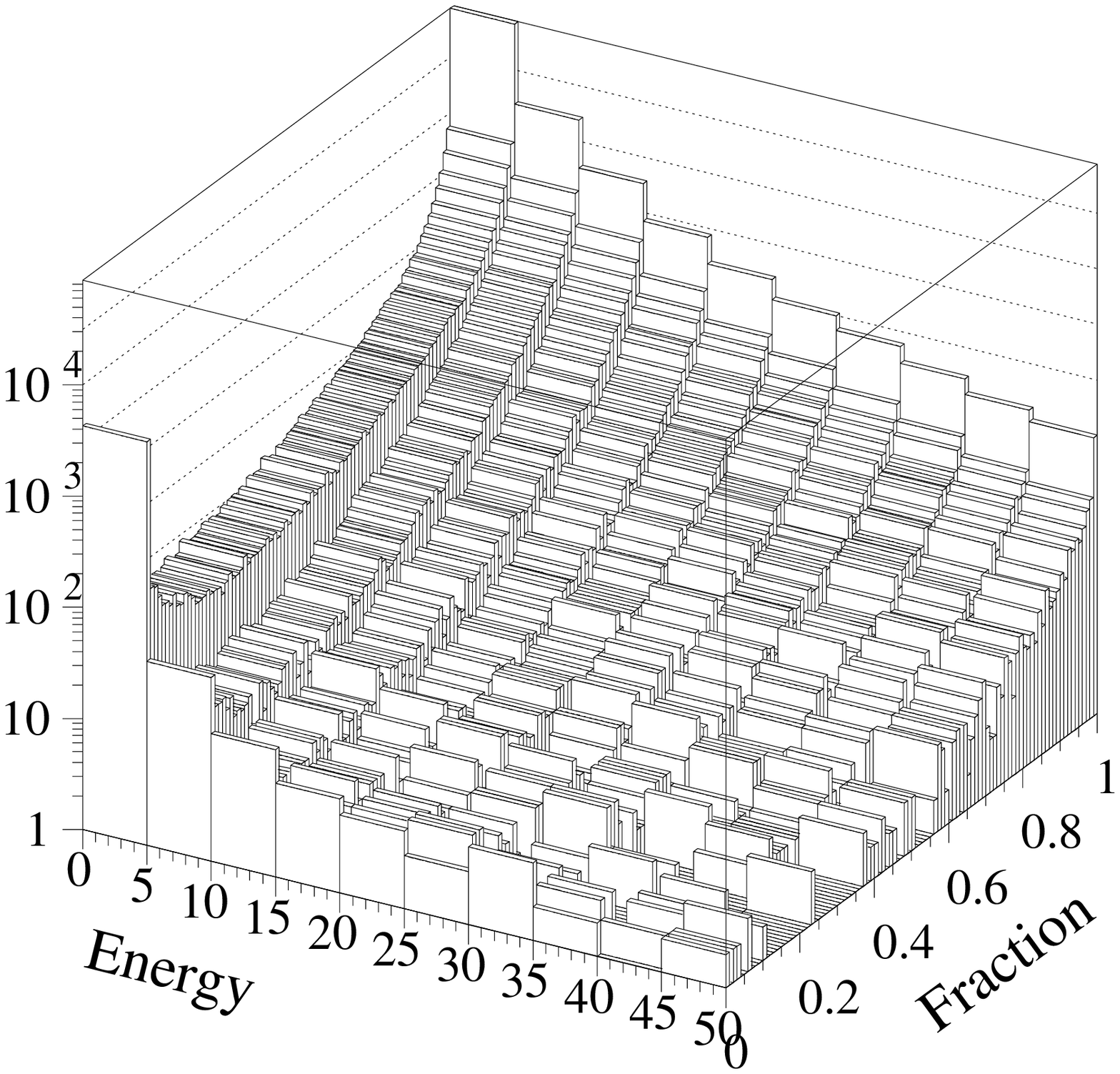}
      \includegraphics[scale=0.32]{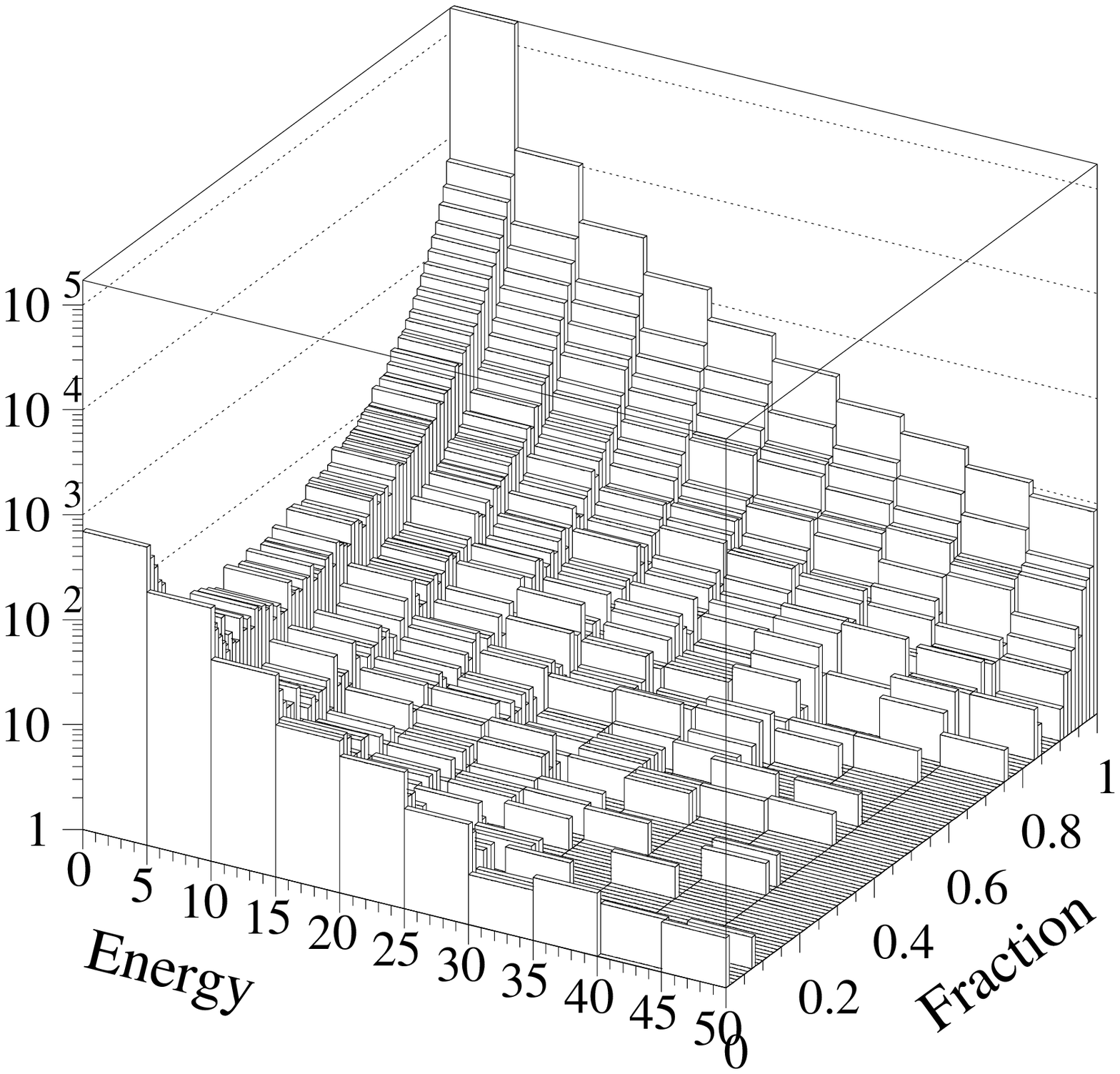}
\caption{Fraction of cluster-energy correctly attributed
 versus cluster energy. The left plot
 shows the situation for charged hadrons (double-counting),
while the right plots shows the situation for photons (energy loss).}\label{Fig:full_frac}
\end{figure}
\begin{figure}[b]
      \includegraphics[scale=0.17]{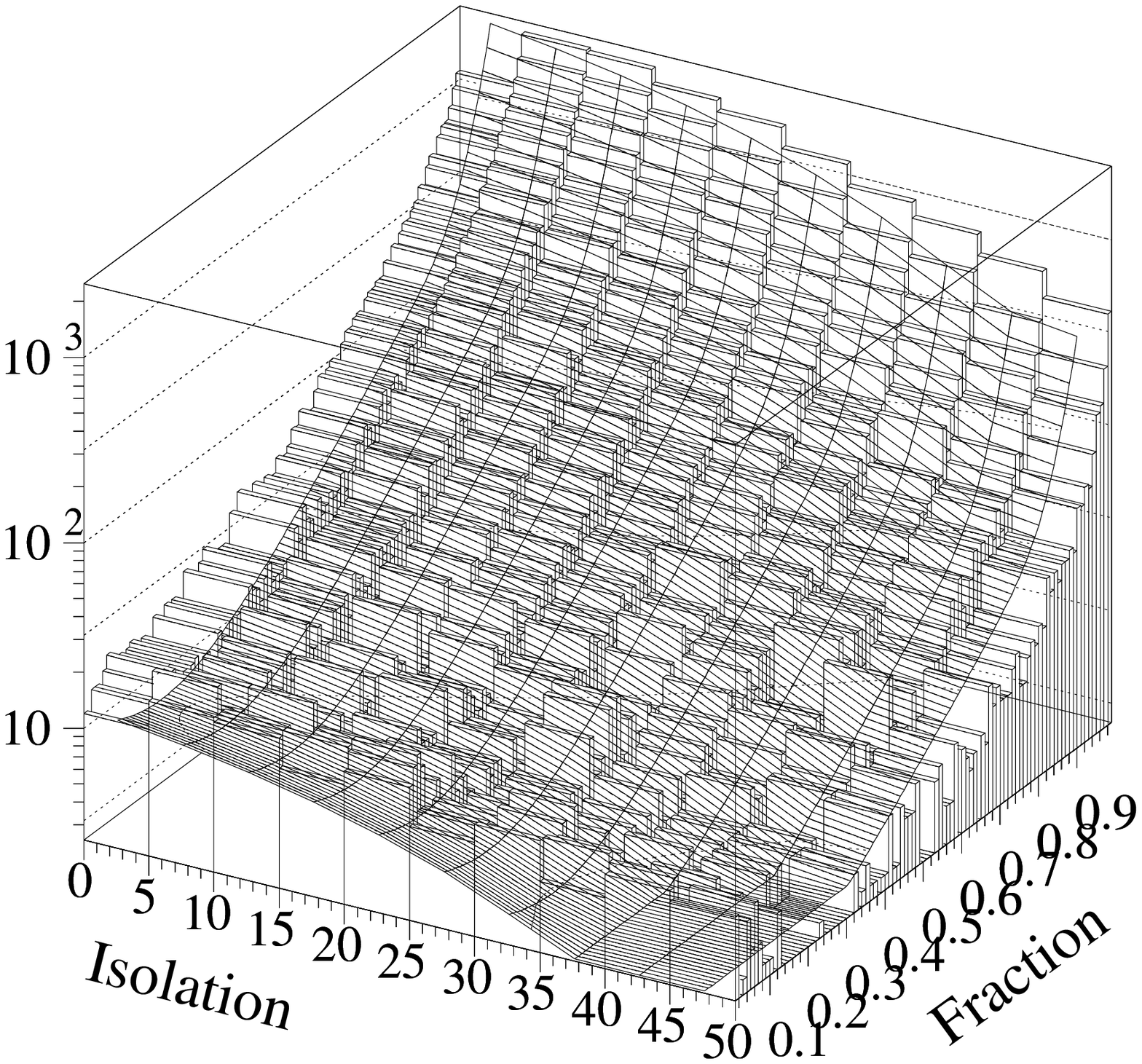}
      \includegraphics[scale=0.17]{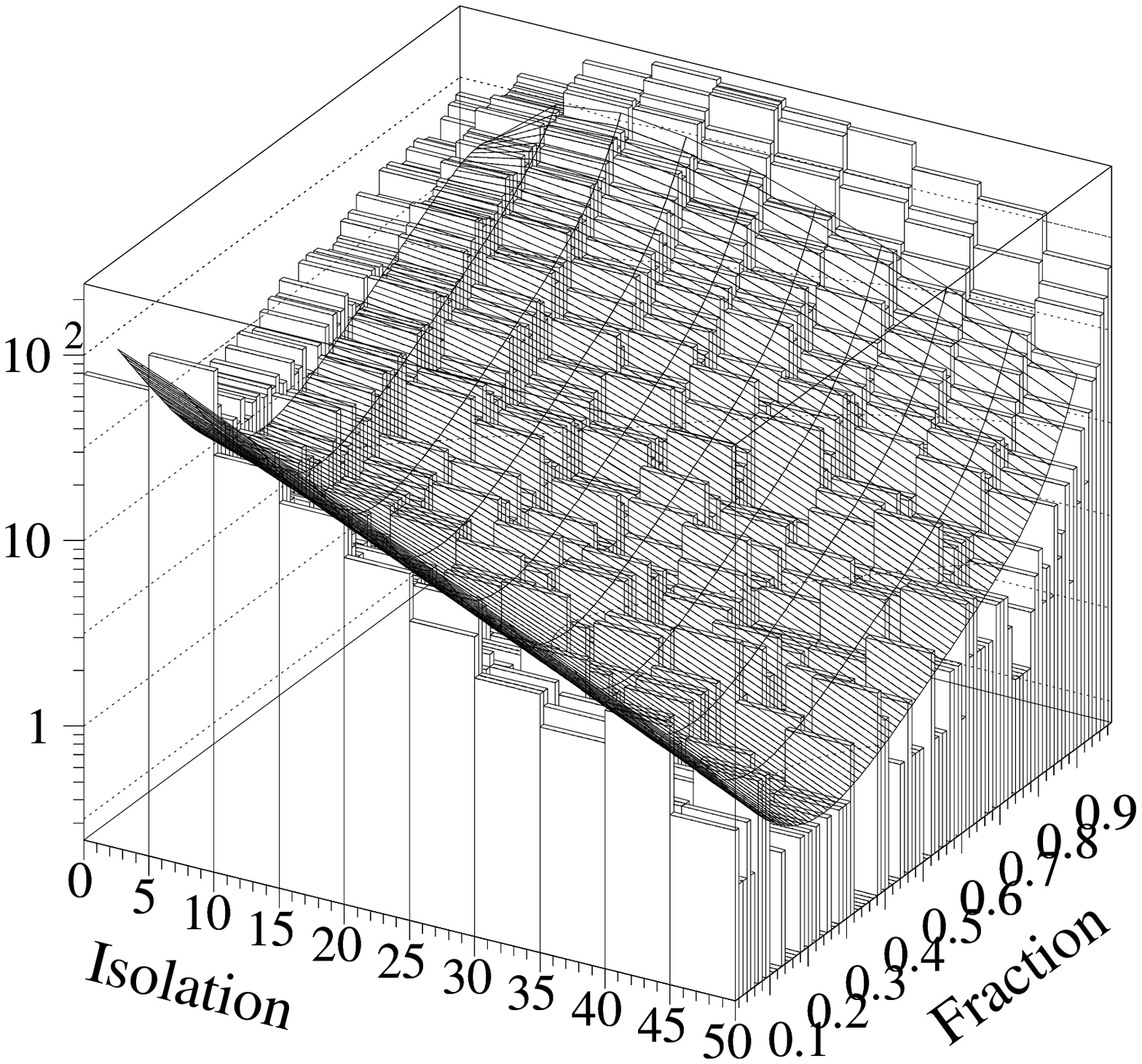}
      \includegraphics[scale=0.17]{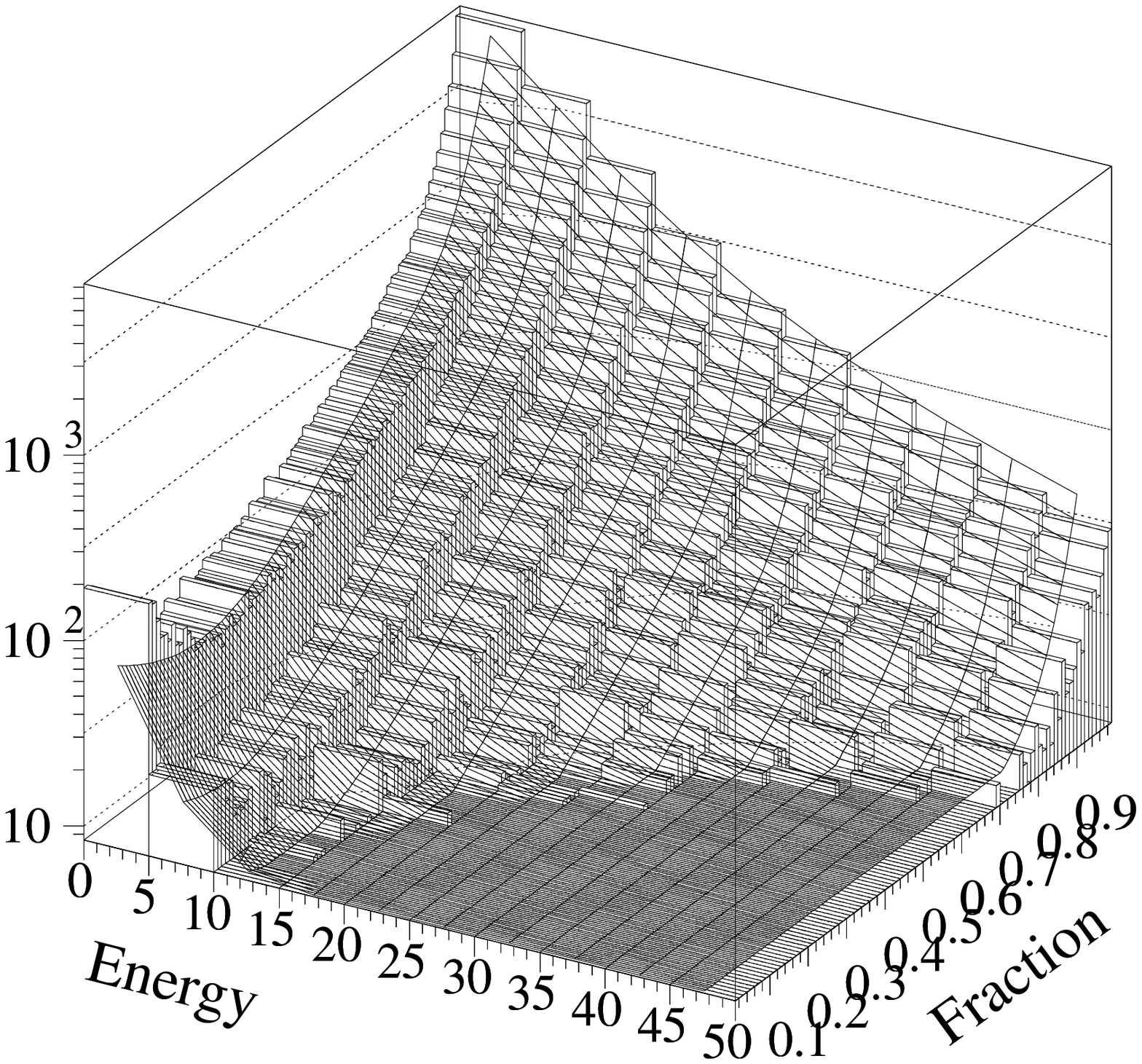}
      \includegraphics[scale=0.17]{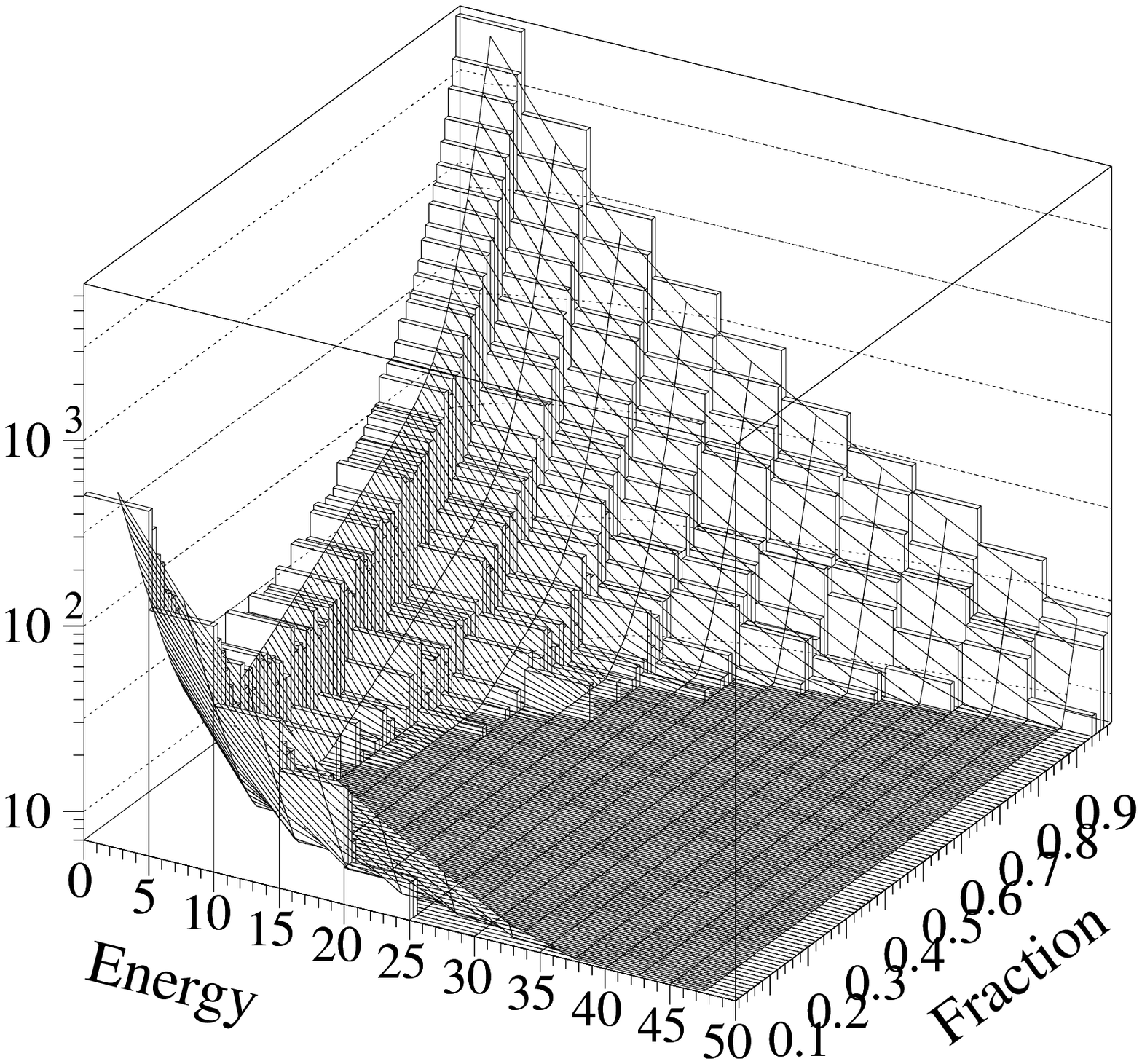}
\caption{Fraction of cluster-energy correctly attributed versus either
isolation (the two plots to the left) or cluster energy (the two plots to the right). 
The plot to the left in each case
 shows the situation for charged hadrons (double-counting),
while the one to the right shows the situation for photons (energy loss).
The histograms shows the observed performance of Pandora, while the
mesh is the final fit. The bins with fraction 0 (complete split) and
1 (no split) are suppressed.}\label{Fig:n01_frac}
\end{figure}
\begin{figure}[t]
      \includegraphics[scale=0.32]{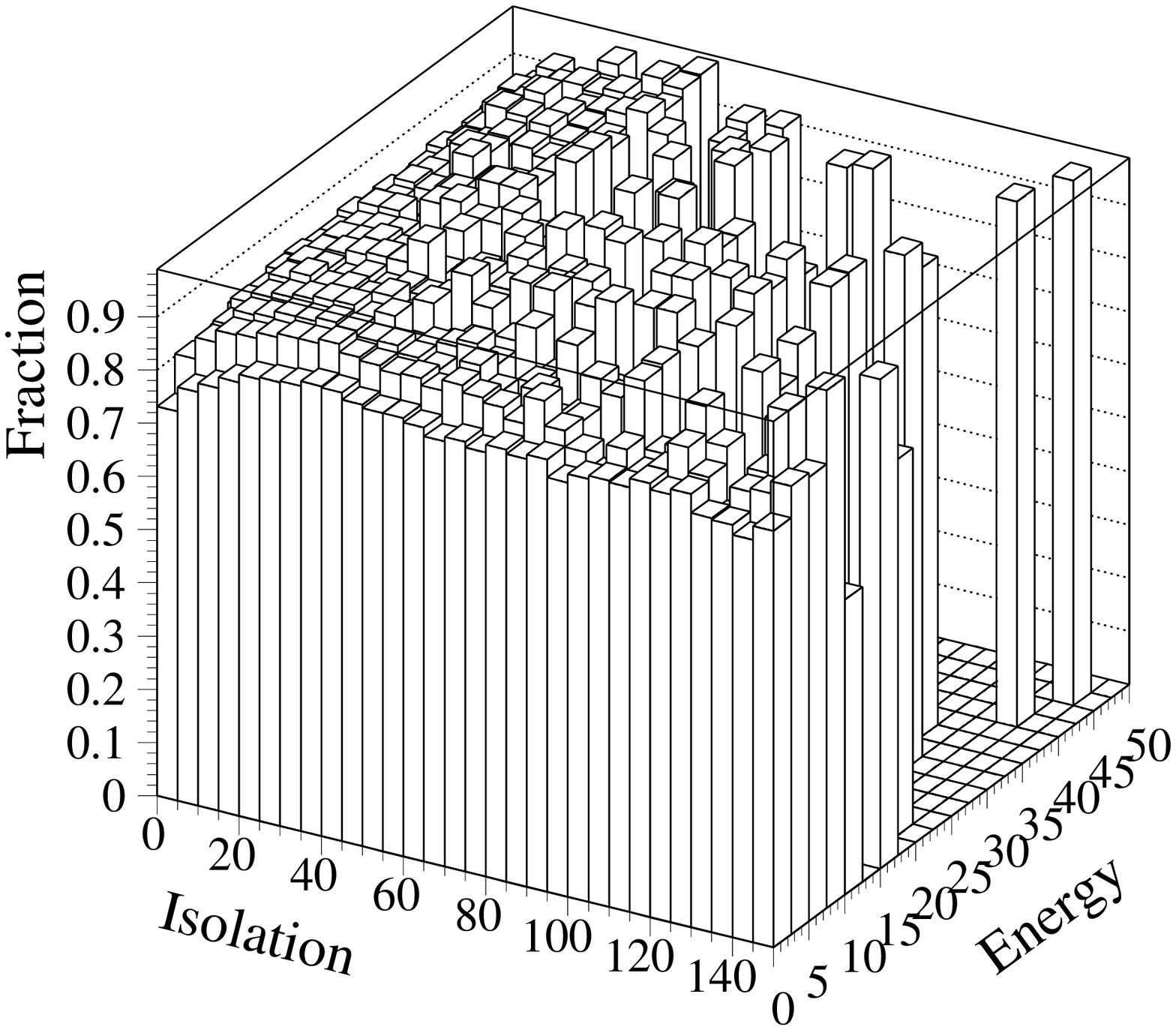}
      \includegraphics[scale=0.32]{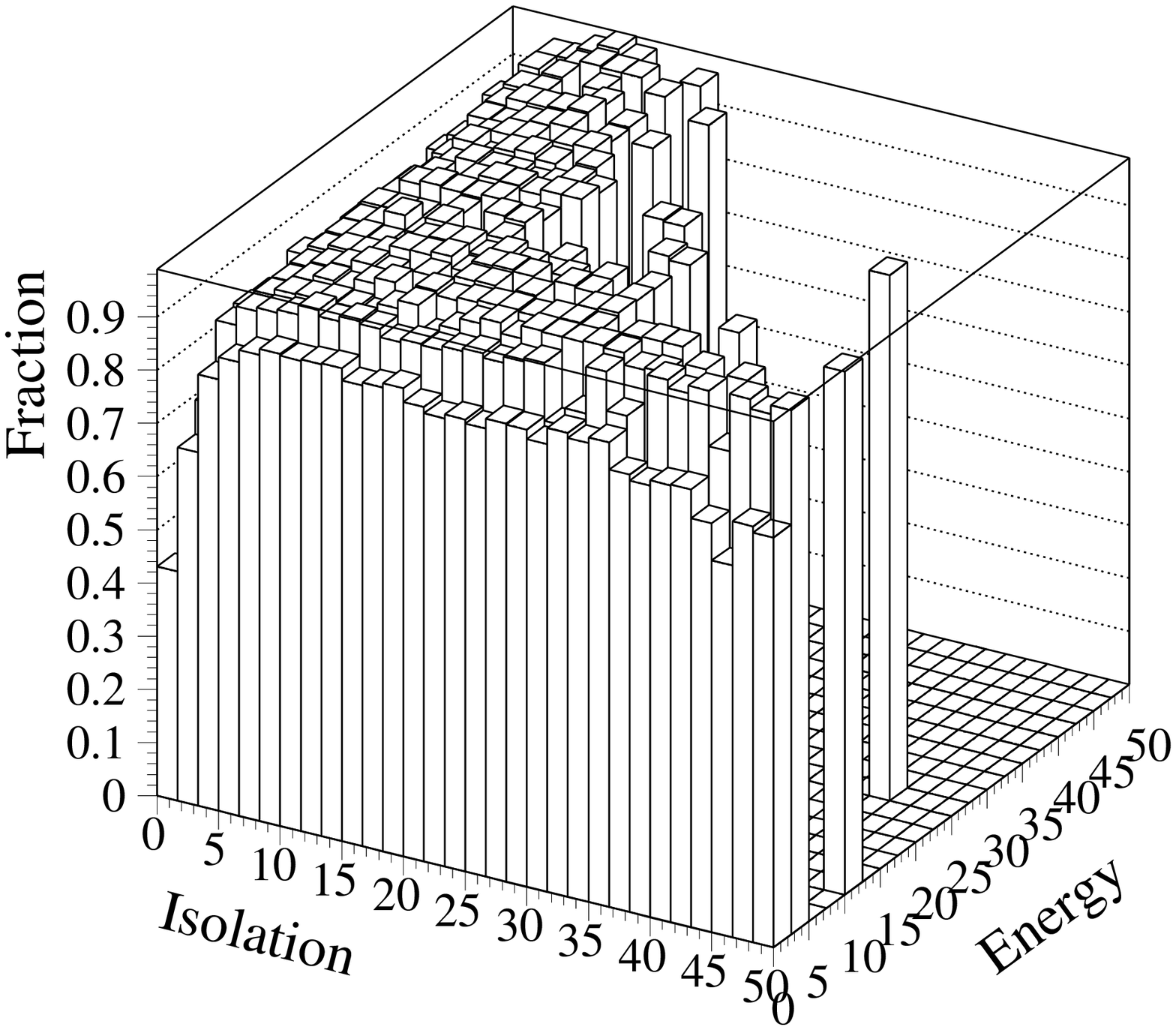}
\caption{The average of the correctly assigned fraction of the cluster energy
versus isolation and cluster energy. Right: the situation for charged hadrons (double-counting).
Left: the situation for photons (energy loss). }\label{Fig:avefrac}
\end{figure}
  \begin{enumerate}[noitemsep]
  \item The probability that a cluster would split: The splitting probability.
  \item In the case the cluster did split: the probability to split off/merge the {\it entire} cluster:
       The complete-split probability.
  \item If the case cluster did split, but not completely: the form of the p.d.f. of the fraction split off:
   The split-fraction.
  \end{enumerate}
One could observe that
  \begin{enumerate}[noitemsep]
 \item The splitting probability depends on the isolation - strongly for energy loss, 
slightly for double-counting -  but 
            can be treated in two energy bins with no
            energy dependence in the bin, as can be seen in Figure \ref{Fig:split}.
 There was also a \%5 over-all dependence on whether the particle was observed in the barrel or end-cap.
  \item  The complete-split probability depends only on the particle's energy, 
  as can be seen in Figure \ref{Fig:full_frac}, by looking at the fraction = 0 bin. 
  \item  The split-fraction depends on both energy and isolation, see Figure \ref{Fig:n01_frac}.
    However, it was also found that the energy and distance
    dependence of the shape could be described by 
    how the average fraction depended on these variables. This dependence is shown
    in Figure \ref{Fig:avefrac}
  \end{enumerate}
\begin{figure}[b]
\includegraphics[scale=0.32]{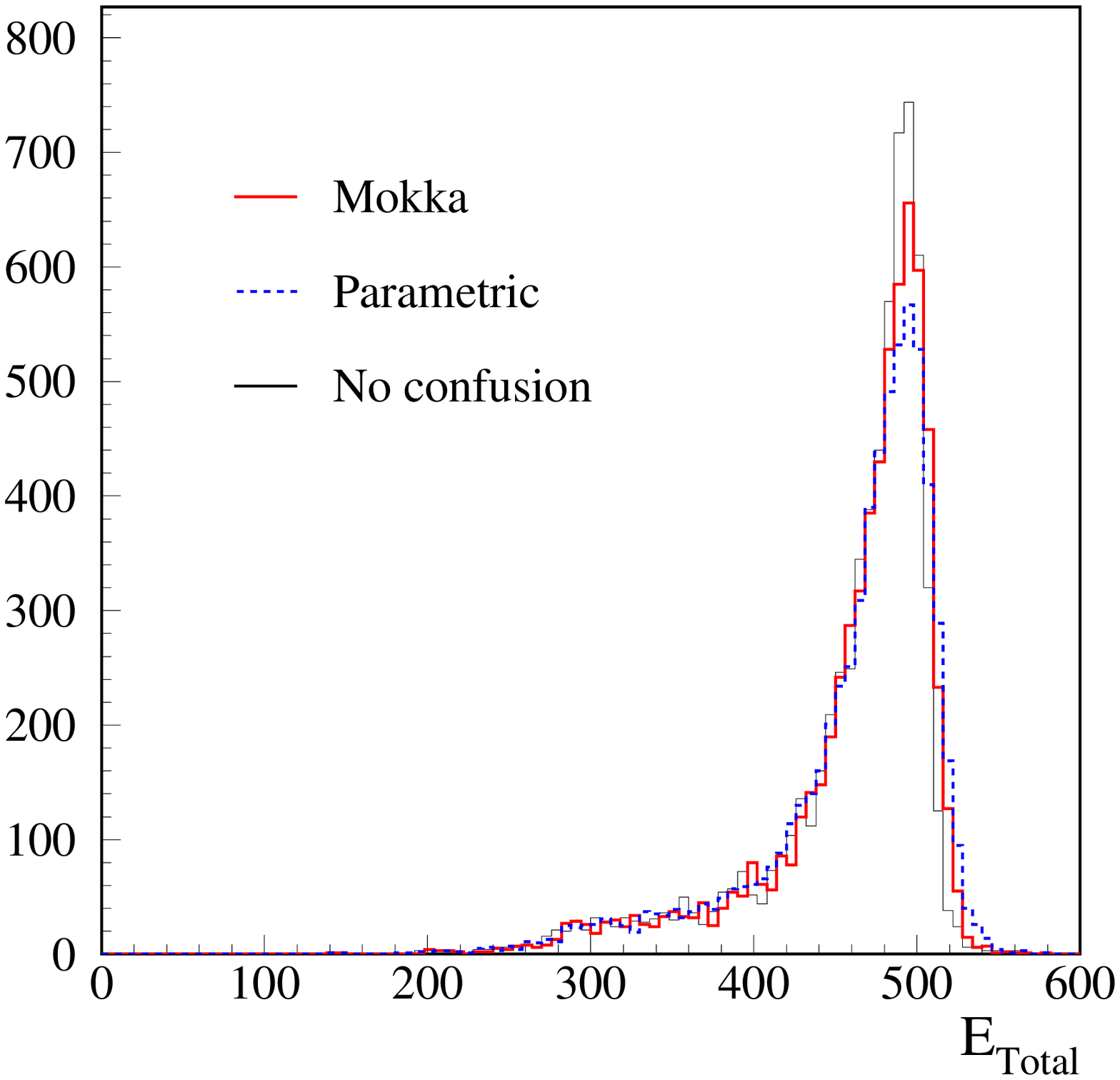}
\includegraphics[scale=0.32]{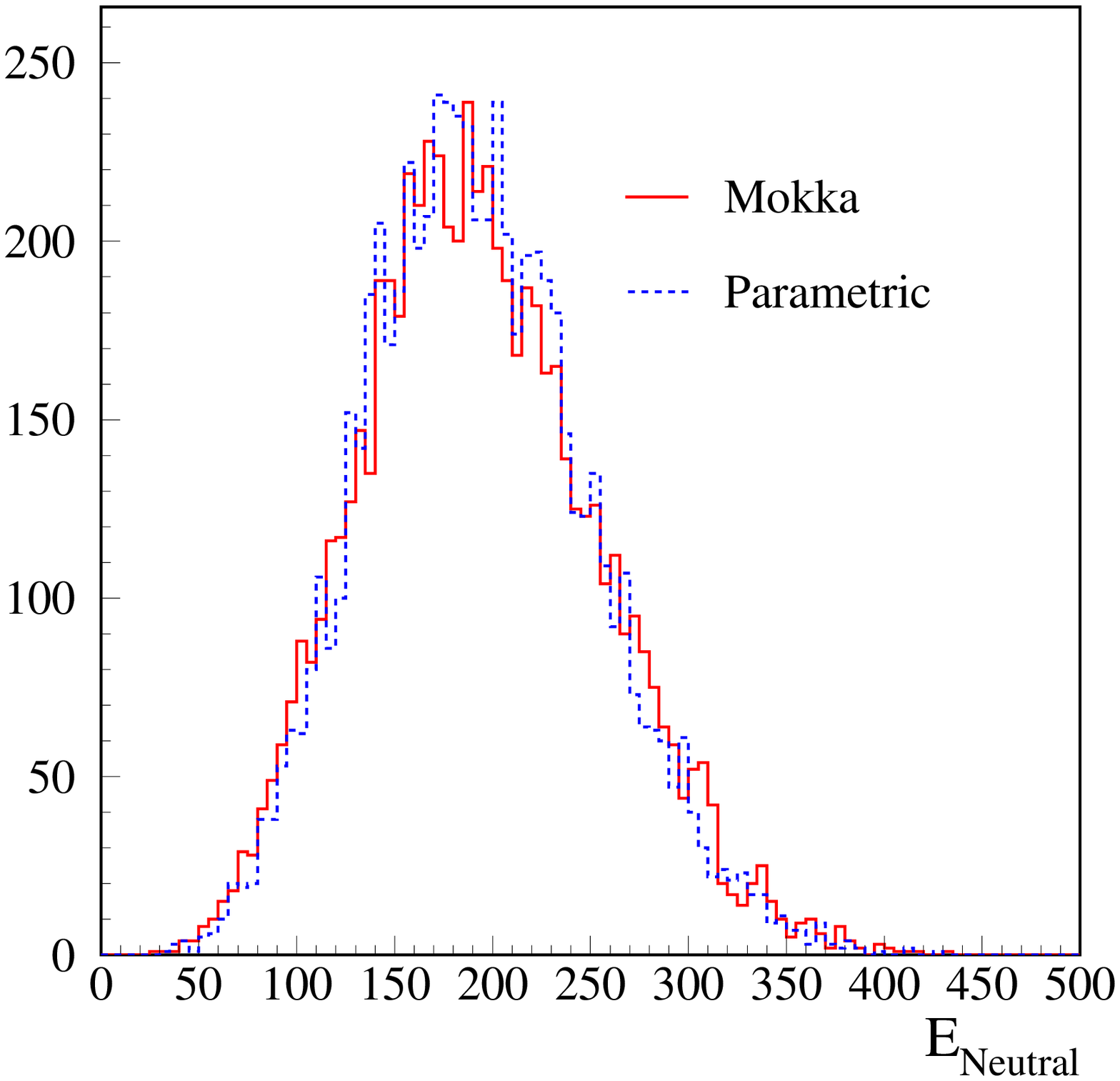}
\caption{Total seen energy (left), and total seen neutral energy (right).
The red line shows the full reconstruction, the blue dashed curve,
the parametric smearing of confusion, and the black solid line the case 
with no confusion.}\label{Fig:etot_n_n}
\end{figure}
All cases (electromagnetic or hadronic cluster - double-counting or energy loss - Barrel or end-cap) 
can be described by the
same functional shapes, only the parameter-values differ between the cases.
The fitted functions could be conveniently chosen as combinations of exponentials and lines.
A total of 28 parameters $\times$ 4 cases (em/had $\times$ double-counting/loss)
are found to be needed.

\begin{figure}[t]
\includegraphics[scale=0.32]{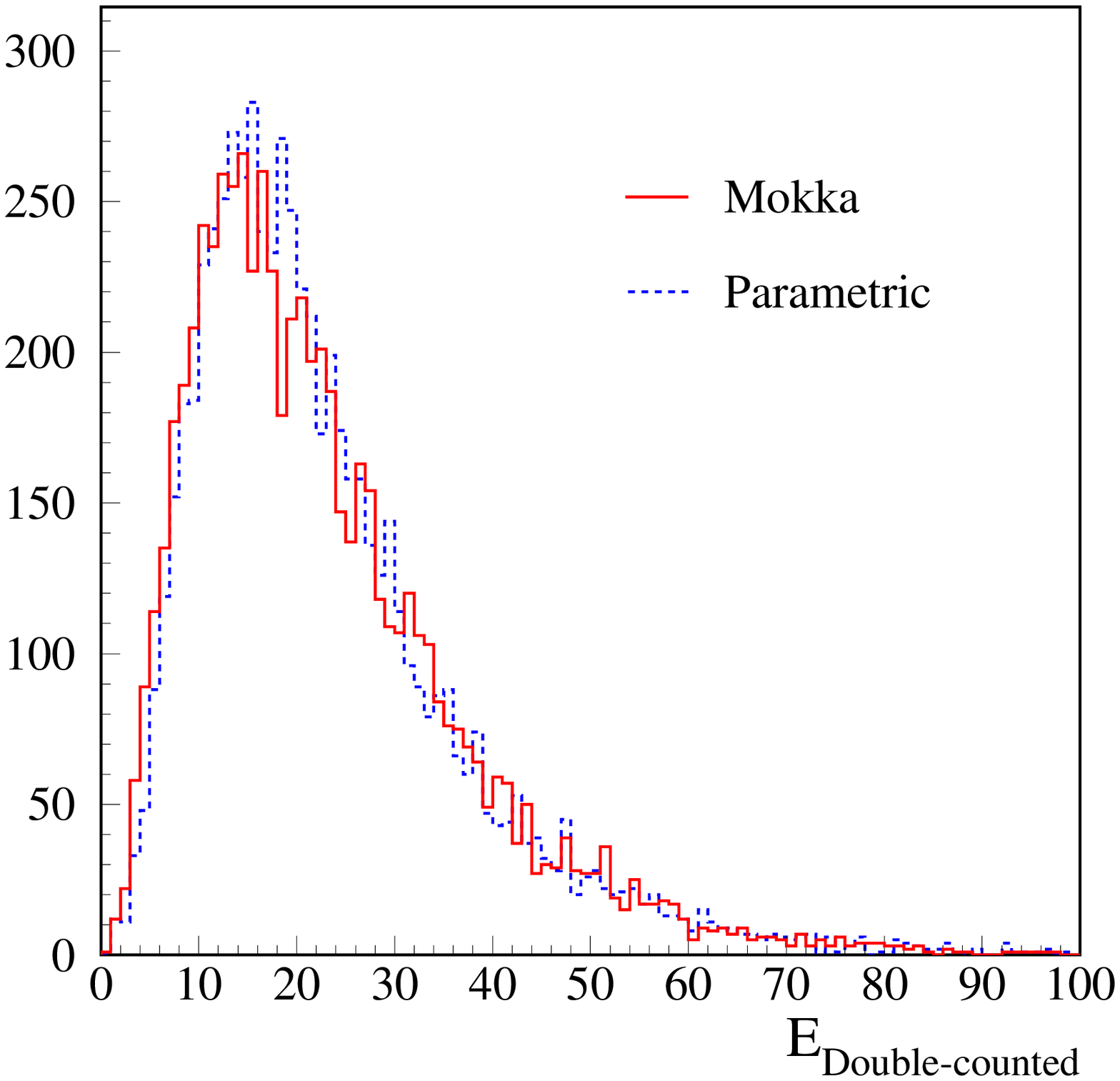}
\includegraphics[scale=0.32]{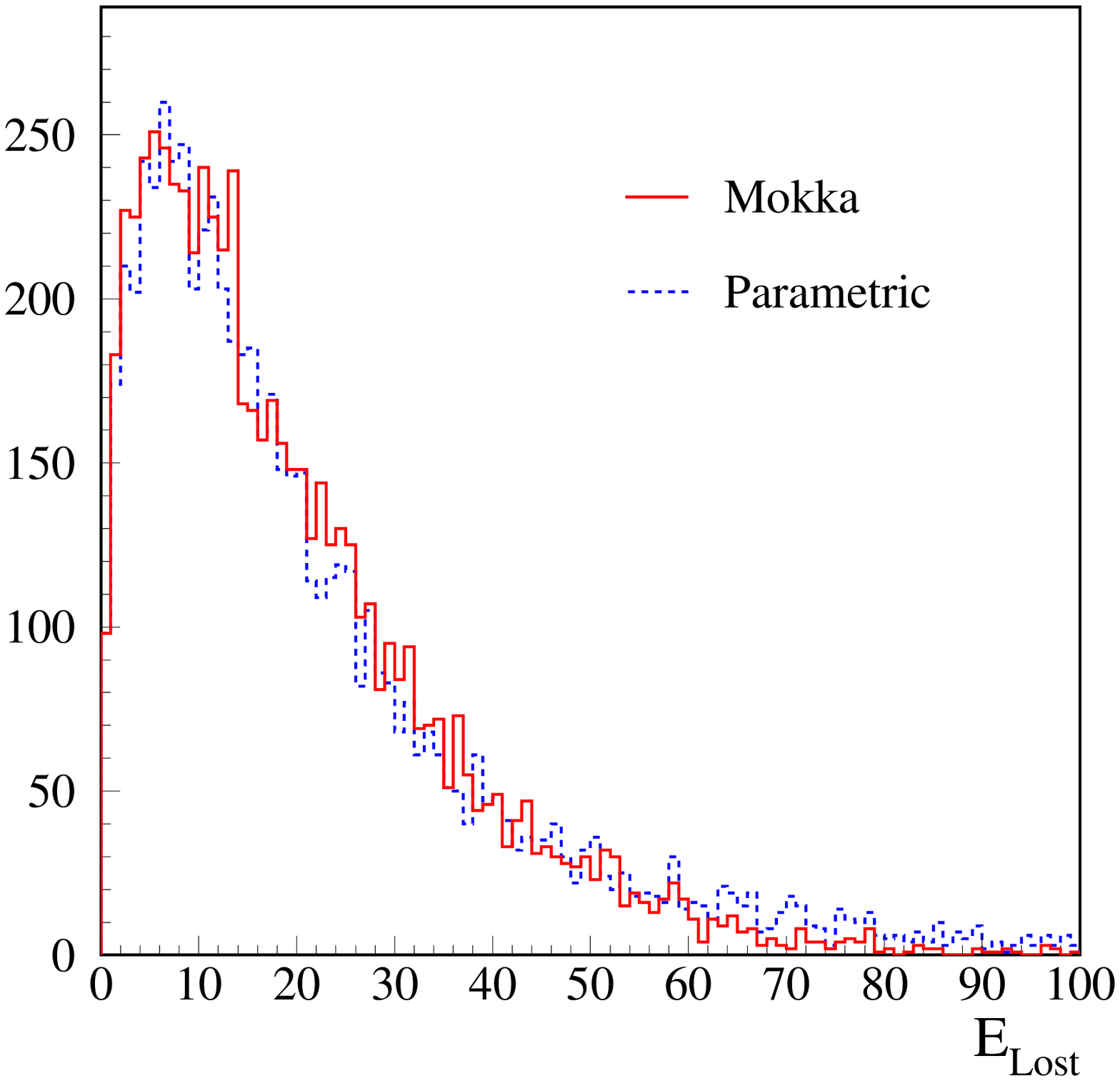}
\caption{Double-counted energy (left) and lost energy (right). The curves have the
same meaning as in Fig \ref{Fig:etot_n_n}.}\label{Fig:edc-elost-best}
\end{figure}

When analysing the  fully simulated and reconstructed sample, the
three fitted functions could be used to simulate double-counting or energy loss 
for each true particle. This parametrically simulated detector-response could then
be compared with the results of the full reconstruction.
A number of global parameters were adjusted to get the
best possible agreement. These parameters include the ratio between 
cluster-energy and track momentum for charged particles and the overall
probability to split clusters.
In Figure \ref{Fig:etot_n_n}, the total seen energy and
the total neutral energy distributions are shown, and  Figure \ref{Fig:edc-elost-best}
shows the
lost and double-counted energy distributions.
One can observe that a quite good agreement was obtained both for the
amount of the two contributors to the confusion (double-counting and
energy loss) and for the global event variables. By studying the
width of the three curves in the total energy figure, it can be noted that
the parametric confusion term is somewhat larger than what the full reconstruction
yields.
Therefore, the SGV with this tuning applied would be somewhat on the pessimistic side,
which is un-usual for fast simulation programs.

\section{Conclusions}

We have pointed out need for fast simulation programs,
both in order to be able to quickly evaluate new theories
confronted with a realistic experimental situation, and
to cope with cases where multi-billion event samples would be
requires {\it viz.} 
large cross-sections ($\gamma\gamma$), or large
                    parameter-spaces in new physics scenarios.
The SGV program was presented, and was shown
to fulfil the requirements emerging from these considerations, both in terms of physics and of computing
performance.
We presented the tracking performance of SGV and found it to be
close to identical to what the full simulation and reconstruction of the ILD
detector yields.
In addition, the way to parametrically incorporate the effects of confusion
between calorimetric clusters was presented. It was shown
that a modest number of parameters were needed to get a result
comparable to the result of full shower development programs.
The procedure was in fact such that the fast simulation result
falls on the somewhat pessimistic side. 
The shower-parametrisation is still work in progress, and would
need future validation on a larger set of physics channels.

\section{Acknowledgements}
The author would like to thank P. Billoir for providing the core code
of the covariance-machine, and F. James for pointing to the method
of Cholesky decomposition. B. Jeffery provided useful help for
the LCIO interface. M. Chera and L. Dai contributed to the
analysis of shower-shapes. A. de Angelis, C. de la Vassi\`ere  and 
M. A. do Vale contributed with various routines, and the comments
from K. Hultqvist, P. Luzniak and V. Adler were very helpful.
Thanks is also due to F. Richard, who initiated the project.
The support of the Deutsche Forschungsgemeinschaft through
the Sonderforschungsbereich 676 (grant SFB 676/2-2010) is acknowledged.

\section{Bibliography}

%% If possible please use BibTeX information as given by INSPIRE
%% to make the citations~\cite{parton_qed} uniform and follow the 
%% examples~\cite{parton_qed,H1,DVCS,pomeron} given below.
%% Note that there is a (non-breaking) space before \verb?\cite?.

% ****************************************************************************
% BIBLIOGRAPHY AREA
% ****************************************************************************

\begin{footnotesize}
% IF YOU DO NOT USE BIBTEX, USE THE FOLLOWING SAMPLE SCHEME FOR THE REFERENCES
% ----------------------------------------------------------------------------

% ----------------------------------------------------------------------------

\end{footnotesize}

\end{document}

%% file: newcom.tex
\def\leqsim{\mathbin{\;\raise1pt\hbox{$<$}\kern-8pt\lower3pt\hbox{$\sim$}\;}}
\def\geqsim{\mathbin{\;\raise1pt\hbox{$>$}\kern-8pt\lower3pt\hbox{$\sim$}\;}}
\def\p#1{\mbox{$ \mbox{\bf p}_1                                         $}}

\newcommand{\eeto}    {\mbox{$ {\, \mathrm e}^+ {\mathrm e}^- \to             $}}

\newcommand{\GeV}     {\mbox{$ {\mathrm{GeV}}                              $}}

\newcommand{\ba}{\begin{array}}
\newcommand{\ea}{\end{array}}
\newcommand{\bc}{\begin{center}}
\newcommand{\ec}{\end{center}}
\newcommand{\be}{\begin{eqnarray}}
\newcommand{\eeq}{\end{eqnarray}}
\newcommand{\bes}{\begin{eqnarray*}}
\newcommand{\ees}{\end{eqnarray*}}
\newcommand{\Kz}{\ifmmode {\rm K^0_s} \else ${\rm K^0_s} $ \fi}
\newcommand{\Zz}{\ifmmode {\rm Z^0} \else ${\rm Z^0 } $ \fi}
\newcommand{\xxbar}{\ifmmode {\rm x\bar{x}} \else ${\rm x\bar{x}} $ \fi}
\newcommand{\rphi}{\ifmmode {\rm R\phi} \else ${\rm R\phi} $ \fi}
\def    \missEt      {\ifmmode{/\mkern-11mu E_t}\else{${/\mkern-11mu E_t}$}\fi}
\def    \missE       {\ifmmode{/\mkern-11mu E}\else{${/\mkern-11mu E}$}\fi}
\def    \missp       {\ifmmode{/\mkern-11mu p}\else{${/\mkern-11mu p}$}\fi}
\def    \misspt      {\ifmmode{/\mkern-11mu p_t}\else{${/\mkern-11mu p_t}$}\fi}